# Optimization of Generalized Unary Coding

Rakshitha Ravula

**Abstract** In this paper, an optimum version of the recently advanced generalized unary coding [9] is proposed. In this method, the block of 1's that identifies the number is allowed to be broken up which extends the count. The result is established by a theorem. The number count is now $n(n-k-1) +1$ rather than the previously described $(n-k)^2-1$.

1. **Introduction**

Unary coding is found in the representation of information in biological systems as in the keeping of time in birdsong [1],[2]. It has also been used applications not only to computer arithmetic [3], neural network training [4], [5] and other coding applications (e.g. [6]). The history of unary coding goes back to the beginning of writing [7],[8].

The main shortcoming of the unary code is its relative inefficiency in representing number count. Recently, a generalized version of the standard unary code was proposed [9] in which instead of a count of n numbers using a n-bit code, the counting was extended to $(n-k)^2-1$, where the number is represented by a string k 1s in the block. In this paper, we present a variant of this generalized coding scheme which extends the count to $n(n-k-1) +1$.

2. **Generalizations of unary coding**

The generalization of unary coding [9] may be done in a variety of manner depending on how the 1s are used in relation to the 0s.

I. **Increasing k**

Increase k until k=n, where n is the size of the block. Table 1 presents the example of n=4 and the resulting count:

Table 1. Increasing k

| n | Code |
|---|------|
| 0 | 0000 |
| 1 | 0001 |
| 2 | 0010 |
| 3 | 0100 |
| 4 | 1000 |
| 5 | 0011 |
| 6 | 0110 |
| 7 | 1100 |
| 8 | 0111 |
| 9 | 1110 |

| 10 | 1111 |

The total count for n bits is 0 to n(n+1)/2. Here n=4, total count is from 0 to 10. The first cycle will count to n, the second to *n−1*, and so on. The total, therefore, is *n + (n − 1) + (n − 2)+ · · · +(n − n + 1) = n2 − (1 + 2 + 3+ · · · +n − 1) =n(n + 1) − n(n + 1)/2 = n(n + 1)/2*.

## II. Fixed k

Here k is fixed and after the k+1 digits, the extra digits are marked by a 1 that is separated from the basic set of k 1s. The separation is first 1 unit, and then it is successively increased. Table 2 illustrates this for n=7 and k=3.

Table 2. Fixed k

| n | Code |
|---|------|
| 0 | 0000000 |
| 1 | 0000111 |
| 2 | 0001110 |
| 3 | 0011100 |
| 4 | 0111000 |
| 5 | 1110000 |
| 6 | 0010111 |
| 7 | 0101110 |
| 8 | 1011100 |
| 9 | 0111001 |
| 10 | 1110010 |
| 11 | 0100111 |
| 12 | 1001110 |
| 13 | 0011101 |
| 14 | 0111010 |
| 15 | 1110100 |

In the fixed k method for n bits where the additional cycles are marked by a distance of 1 and more in succession, the total count is 0 through $(n-k)^2-1$. The count is *n − k+1* in the first cycle and each of the subsequent cycles. The total number of cycles possible is *n − k − 1*. Therefore, the total count is *(n − k + 1)(n − k − 1) = (n − k)2 − 1*.

Here for n=7, k=3 total count is from 0 to 0 to 15.

### 3. Proposed Method

Here k is fixed and for every value of s a bit is moved towards left. n is the total number of bits including k 1's. For every cycle i.e. after every (s-1) digits starting from s=2, a 1 is appended to n and this is separated from the basic k 1's by a zero. For additional cycles the number of zeros between the appended 1 and basic set k 1's is increased by 1.
Here is an example for n=8, k=3

Table 3: Fixed K moving 1 bit at a time

| n | Code |
|---|---|
| 0 | 00000000 |
| 1 | 00000111 |
| 2 | 00001110 |
| 3 | 00011100 |
| 4 | 00111000 |
| 5 | 01110000 |
| 6 | 11100000 |
| 7 | 11000001 |
| 8 | 10000011 |
| 9 | 00010111 |
| 10 | 00101110 |
| 11 | 01011100 |
| 12 | 10111000 |
| 13 | 01110001 |
| 14 | 11100010 |
| 15 | 11000101 |
| 16 | 10001011 |
| 17 | 00100111 |
| 18 | 01001110 |
| 19 | 10011100 |
| 20 | 00111001 |
| 21 | 01110010 |
| 22 | 11100100 |
| 23 | 11001001 |
| 24 | 10010011 |
| 25 | 01000111 |
| 26 | 10001110 |
| 27 | 00011101 |
| 28 | 00111010 |
| 29 | 01110100 |
| 30 | 11101000 |
| 31 | 11010001 |
| 32 | 10100011 |
| 33 | 10000111 |

**Theorem**: In fixed k method moving 1 bit each time for n bits, the total count is 0 through n(n-k-1) +1.

**Proof**: There are k 1s in the n bit long sequence and hence there are (n-k) 0s. The unique sequences that are formed without considering the shifts are (n-k-1). Therefore, the total count in this process will be n(n-k-1) because each of these can be shifted n times. The last count will be the termination of the process. Hence, total count is from 0 to n(n-k-1)+1.

4. **Algorithm for Coding and Decoding**

We now present methods for coding and decoding of the proposed scheme.
**Coding:**

i. Inputs given n, k. Calculate n-1
ii. For s=0, n=00000000 (always). For n=1, append basic set of k 1's to the right.
iii. Shift each bit left for every increase in s value. This cycle repeats for every (n-1) times.
iv. After s= 2n+1, the number of 0's between the appended 1 and basic set of k 1's is incremented.
v. This process continues till the final count is n(n-k-1) +1.

**Decoding:**

i. Calculate n, k.
ii. P denotes the multiple of n. P= {1, …. , (n- k)}.
iii. For every s=Pn+1, the number of 0's between the appended 1 and basic set of k 1's is incremented.

Here consider an example n=8, k=4. To calculate 13:

i. Find 1,9, 17,25
ii. 13 lies between 9 and 17.
iii. 13-9=4, so move first 4 bits to right and keep the rest of the bits as it is.

5. **Analysis of different cases with fixed k: k=3, n=8, $n_1$=1**

We now present results on the distance between different codewords which could be relevant in error situations. The distance function varies in a zig-zag manner for both the previous scheme as well as the new scheme proposed by us as shown by Figure 1.

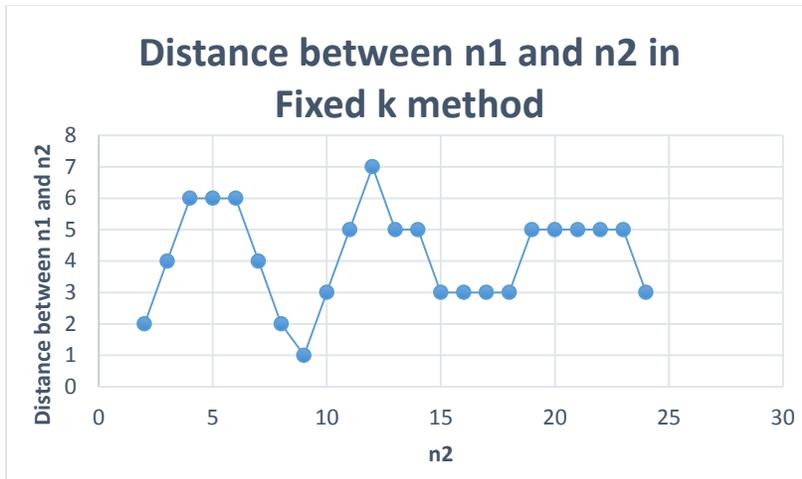

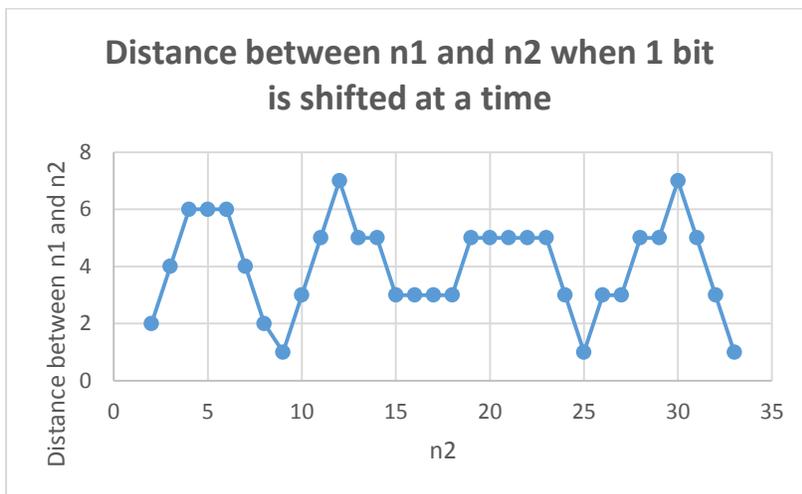

Fig 1. Fixed k, on X axis: $N_2$; on Y-axis: Distance between n1 and n2

From the graph we observe the proposed scheme in this paper is more efficient as it has higher count than the previous method. The count in the first method is $(n-k)^2-1$ and the count in the second method is $n(n-k-1)+1$. For fixed k=3 and varying values of n the variation in the count is represented clearly in the graph below:

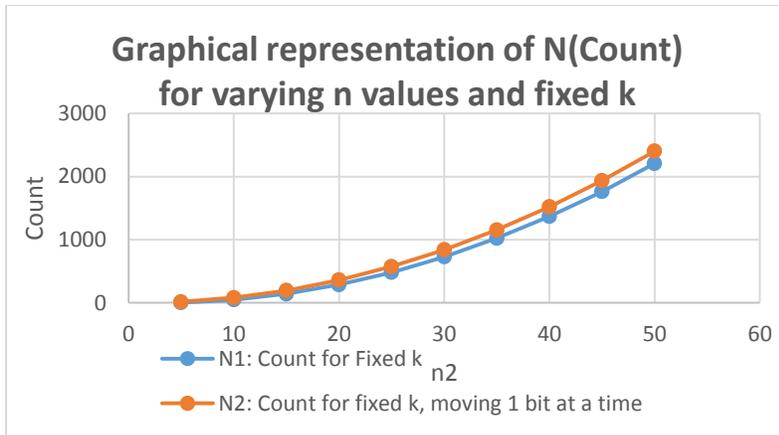

Fig 2: Graphical representation of count for varying n values and fixed k

### 6. Conclusions

An optimum version of the recently advanced generalized unary coding is proposed which extends the capacity to count beyond the result of the previous method. In this method, the block of 1's that identifies the number is allowed to be broken up which extends the count. The result is established by a theorem. The number count is increased to n(n-k-1) +1 rather than the previously described (n-k)$^2$-1.

7. S. Kak, A frequency analysis of the Indus script. Cryptologia 12: 129-143, 1988.
   S. Kak, Indus and Brahmi: Further connections. Cryptologia 14: 169-183, 1990.
   S. Kak, History of Physical and Chemical Thought in India. https://subhask.okstate.edu/sites/default/files/HistoryPhysicalChemicalThought.pdf

8. G. Ifrah, The Universal History of Numbers. John Wiley, 2000.

9. S. Kak, Generalized unary coding. Circuits, Systems and Signal Processing 36: 1419-1426, 2016.